# Double Percolation Transition in Superconductor/Ferromagnet Nanocomposites

Xiangdong Liu, Raghava P. Panguluri, Zhi-Feng Huang, and Boris Nadgorny

*Department of Physics and Astronomy,*

*Wayne State University, Detroit, MI 48201*

(Dated: May 26, 2009)

## Abstract

A double percolation transition is identified in a binary network composed of nanoparticles of $MgB_2$ superconductor and $CrO_2$ half-metallic ferromagnet. Anomalously high-resistance or insulating state, as compared to the conducting or superconducting states in single-component systems of either constituent, is observed between two distinct percolation thresholds. This double percolation effect, which is especially pronounced at liquid helium temperatures, is controlled by composite volume fraction and originates from the suppressed interface conduction and tunneling between nanoparticles of different species. We investigate the scaling behavior near both percolation thresholds, and determine the distinct critical exponents associated with two different types of transitions.

Percolation effects have been widely investigated in a variety of physical systems [1, 2], with applications ranging from ferromagnetic or superconducting solid materials, such as composites [3, 4] or polycrystalline samples [5], to soft matters, such as polymer blends [6]. In addition to the fundamental importance of understanding phase transitions, the corresponding mechanical, electronic, transport, and magnetic properties of percolating networks are often strongly affected by the system percolation. A typical example would be spintronics applications, where percolation is often associated with large magnetoresistance [3] and may play an important role in the physics of colossal magnetoresistive oxides [7]. In most model systems that have been studied extensively, only one percolative network is considered, with the corresponding percolation theory well established [1, 2, 4, 8]. The problem of electrical conduction and transport in such networks has been of particular interest. This classical percolation problem is usually featured by a conductor-insulator (or conductor-superconductor) transition in a two-component composite, where one of the constituents provides a conducting (superconducting) path through the insulating (normal) matrix [1, 2, 4, 5]. In this case the percolation threshold is uniquely determined by a single conducting (superconducting) constituent. It would be of broader interest, however, to investigate percolation effects in a more complex system, where multiple percolative networks can potentially be formed, as often encountered in real multicomponent materials. Since a general case with arbitrary coupling and variable interaction between different percolative media may be prohibitively complicated, it is instructive to first construct and study a special, well controlled case where different percolation networks can be identified independently [9].

In this Letter we report a new type of double percolation transition, observed in a binary system where both constituents can form independent percolative networks. A remarkable feature of this system is the existence of two separate thresholds, each corresponding to conductor-insulator or superconductor-insulator transition, but in two different media that are conducting or superconducting on their own in a single-component state. The system we study is based on cold pressed powders of $CrO_2$ and $MgB_2$ mixture, with the percolation path arising through contacts between individual nanoparticles. The two materials are well suited for this experiment: $CrO_2$ is a half-metallic ferromagnet where the conduction band is fully spin polarized [12–16], whereas $MgB_2$ is a superconductor with a relatively high bulk critical temperature of 39K. We find that the interface conduction between $CrO_2$ and $MgB_2$ particles is strongly suppressed compared to that of $CrO_2/CrO_2$ or $MgB_2/MgB_2$ interfaces, especially below the superconducting transition [17]. The percolative networks from the two constituents can then be effectively separated, yielding two distinct percolation thresholds. We can thus identify an anomalous *conductor-insulator-superconductor* transition which is attributed solely to the double percolation phenomenon. Our study shows that such double-percolating system is essential for the crossover between conducting/insulating/superconducting states as controlled by volume fraction of the composite. While here we refer specifically to $CrO_2$ and $MgB_2$ composites, our model is applicable to other ferromagnet/superconductor systems, or, more generally to two-component systems with two distinct percolation thresholds that do not overlap.

We study a wide range of such composite powders, where a weight fraction $x$ of $CrO_2$ was mixed with the corresponding fraction $1 - x$ of $MgB_2$. This mixture of commercial $CrO_2$ and $MgB_2$ powders was ground for one hour in a controlled humidity environment, with the humidity not exceeding 30%. Discs with a diameter of 5 mm and thickness of 1-2 mm were cold pressed at a pressure of 10 GPa. Samples with 23 different compositions with $x$ varying from 0 to 1 were prepared; at least two samples were made with the same nominal composition. Immediately after fabrication, the samples were mounted in the Physical Properties Measurement System (PPMS) for four point transport measurements. Both particle size and average particle distribution were monitored by scanning electron microscopy (SEM).

In Fig. 1a we show a typical particle distribution in a composite sample with $x = 0.4$, as obtained from SEM Energy Dispersive Spectrometry (EDS) mapping of elemental Mg and Cr. The SEM images of the original $CrO_2$ and $MgB_2$ powders are given in Figs. 1b and 1c respectively. The typical length of rectangular $CrO_2$ particles is about 300 nm and the width about 40 nm. The $MgB_2$ particles are polydispersed, with approximately spherical shape and the average size of about 500 nm. The typical packing densities are about 46% for $CrO_2$ and 62% for $MgB_2$.

The electrical resistance for the samples of various composition has been measured at different temperatures; see Fig. 2. In the inset we show the resistance of the samples prepared in the same fashion from pure $MgB_2$ (which has a superconducting transition of around 37 K), and pure $CrO_2$ whose resistance is fairly flat except for the lowest temperatures, where the activation type behavior is observed, due to the presence of a nm thick $Cr_2O_3$ insulating layer at the surface of $CrO_2$ [18] . The resistance of the composite samples is low for pure $MgB_2$, $x = 0$, and pure $CrO_2$, $x = 1$, and attains maximum values at fairly narrow intermedi-ate composition range $0.35 < x < 0.5$, with the low-temperature peak value of the resistance about three orders of magnitude higher than that of pure $CrO_2$ sample. We attribute this behavior to a double percolation effect, where two separate percolative networks with conductive or superconductive paths develop at different thresholds $x_c$ for the two constituents: $x_c^{CrO_2} \simeq 0.5$ and $x_c^{MgB_2} \simeq 0.65$ (which corresponds to 65% weight fraction of $MgB_2$). The corresponding concentration thresholds $p_c$ (for volume fraction $p$) is not symmetric, i.e., the values of $p_c^{CrO_2}$ ($\simeq 0.34$) and $1 - p_c^{MgB_2}$ ($\simeq 1 - 0.78 = 0.22$) are not the same, due to the different sizes and shapes of the constituent particles, and also the different nature of the percolation transition. The threshold is also significantly sharper at low temperatures on the $MgB_2$ side, as the percolative path represents not just an insulator-conductor, but an insulator-superconductor transition.

We expect these surprising experimental results to be closely connected to the anomalously high resistance of the $MgB_2/CrO_2$ interface and hence be understood via the two percolation clusters formed separately by the $MgB_2$ and $CrO_2$ nanocomponents. The anomalous behavior at such heterogeneous interface is verified by independent measurements of the resistance of a sandwich structure between two cold-pressed layers composed of pure $MgB_2$ and pure $CrO_2$ nanoparticles respectively, otherwise prepared under the same conditions and in the same geometry as all the other samples. The resistance of such a structure turns out to be about two orders of magnitude higher than the resistance of pure $CrO_2$ layer (see the insets of Fig. 2, where the resistance of pure $MgB_2$ is negligible in comparison). While the microscopic structure of the interface may be fairly complicated, and we can not assume that it is simply a single $MgB_2/CrO_2$ layer, these results clearly indicate that the $MgB_2/CrO_2$ interface resistance is orders of magnitude larger than that of $CrO_2/CrO_2$ or $MgB_2/MgB_2$ interfaces in pure samples, particularly at low temperatures below the superconducting transition.

It can thus be anticipated that the path involving most heterogeneous interface regions would lead to the highest resistance. Note that such interfaces correspond to those between $MgB_2$ and $CrO_2$ clusters, the perimeter of either of which can be estimated as [1] $t_S \sim S(1-p)/p + c_0 S^\zeta$ while close to a percolation transition, where $c_0$ is a constant, the critical exponent $\zeta = 1$ (for $p < p_c$) or $1 - 1/d$ (for $p > p_c$ in $d$ dimensions), and $S$ is the cluster size (or mass) which scales as $(p - p_c)^{-1/\sigma}$ with $\sigma$ another critical exponent, for the "critical" clusters dominating the system properties. Thus maximum cluster perimeter (and hence maximum heterogeneous interface regions) will be reached when $p \to p_c$. For the current binary composite where both conducting constituents (MgB$_2$ and CrO$_2$) can percolate through the system, the new effect of high sample resistance can only occur before the appearance of a percolation cluster (which would then be either conductive or superconductive), but close enough to both of the percolation transitions to maximize the MgB$_2$/CrO$_2$ heterogeneous interfaces. This will lead to a maximum resistance (or insulating) state in between the two concentration thresholds, as demonstrated in Fig. 2.

To further illustrate the existence of such double percolation effect, we examine in Fig. 3 the scaling behavior of the resistance (or dc conductivity) of the binary composite near the two percolation thresholds. Based on the standard percolation theory [1, 2, 4], a power law behavior for resistance $R$ is expected while approaching each transition threshold $p_c^{(i)}$ (with ($i$) referring to CrO$_2$ or MgB$_2$ network), i.e.,

$$R \sim \left|p - p_c^{(i)}\right|^{-\mu}, \tag{1}$$

where the critical exponent $\mu$ depends on the dimensionality and the intrinsic conducting property of the system: e.g., in three-dimensional (3D) conductive networks, $\mu \simeq 2$ for lattice percolation and $\mu \simeq 2.38$ for continuum percolation (the Swiss-cheese model), while in 3D lattice of superconductivity $\mu$ (usually denoted as $s$) is about 0.74 for a conductor-superconductor network [1, 2]. Such scaling behavior has been confirmed in our experimental results for both MgB$_2$- and CrO$_2$-dominated networks, as shown in Fig. 3a and 3b respectively (with the use of $p_c^{\mathrm{CrO_2}} = 0.34$ and $p_c^{\mathrm{MgB_2}} = 0.78$). As we approach the threshold on the CrO$_2$ side of the insulator-conductor transition (i.e., near $x = 0.5$ ($p = 0.34$) of Fig. 3), the data rescaling presented in Fig. 3a yields the exponent $\mu = 2.16 \pm 0.07$, which is between the theoretical values of 2 (for 3D lattice) and 2.38 (for 3D continuum percolation) as given above. The experimental value for the threshold of the CrO$_2$ network ($p_c^{\mathrm{CrO_2}} \simeq 0.34$) is also close to the theoretical threshold of 3D site percolation in simple cubic lattice ($p_c \simeq 0.31$) or that of 3D continuum percolation model (the Swiss-cheese model with uniform spheres; $p_c \simeq 0.28$) [1, 2]. On the other hand, for the MgB$_2$ side of the superconducting network at low temperatures, a different critical exponent of $s = 1.37 \pm 0.05$ is identified from Fig. 3b. As can be expected, this exponent deviates from the known theoretical result ($s \simeq 0.74$) for conductor-superconductor networks, due to the new type of insulator-superconductor transition taking place in this system. For the MgB$_2$ network, our experimental threshold of $p_c^{\text{MgB}_2} = 0.78$ could only be compared to the known results of superconductor/conductor thresholds of, e.g., MgB$_2$ and YBa$_2$Cu$_3$O$_7$, which are in the range of $0.2 - 0.6$ [19, 20]. While percolation thresholds are known to be dependent on the geometric irregularity and polydispersity of the constituent particles and on the fabrication conditions, in our case the influence from the strongly interactive ferromagnetic media (CrO$_2$) might also affect the results, shifting the percolation threshold to a higher value.

While quantitative results of percolation thresholds, scaling exponents, and sample resistance are usually quite sensitive to the system microstructure, such as the size, shape and the distribution of the particles, we expect the observed phenomenon of conductor-insulator-superconductor transition to be robust, as long as the double-percolation thresholds for the two media, I and II, are distinct and $p_c^{\text{I}} + p_c^{\text{II}} > 1$. This latter condition implies that there is a region between the two thresholds where neither constituent percolates, resulting in two separate phase boundaries and hence optimal area of heterogeneous particle interfaces which lead to the double percolation effect described above. Additionally, the nanocrystalline nature of the components is expected to play an important role in the appearance of the observed *insulating* state in the *conducting* or *superconducting* background media. As discussed above, the anomalous conducting transition can be attributed to the interface effects of individual grains of MgB$_2$ and CrO$_2$. Such effects are known to be of increasing importance for gradually decreasing grain size, particularly at the nanoscale where the thickness of the interface region may approach the individual particle size. In this case the influence of interface tunneling, which is expected to be the main mechanism underlying the anomalous interface conducting behavior observed in Fig. 2, will be dominant, especially at low temperatures. This interface-controlled conductance would be analogous to that seen in other percolative systems of nanocrystalline composites, such as Li$_2$O(conductor)/B$_2$O$_3$(insulator) composites [21, 22], though the opposite effect of conductivity enhancement at interfaces was reported there. Further systematic studies, both experimental and theoretical, are needed to identify the material parameters and various processing conditions that give rise to system-specific percolation patterns.

In conclusion, our study of transport properties of ferromagnetic half-metal (CrO$_2$) and superconductor (MgB$_2$) compacted powder samples reveals the presence of double percolative transitions. Both of these transitions obey the conventional scaling laws of percolation – but are characterized by different critical exponents and thresholds, thus leading to an anomalous conductor-insulator-superconductor crossover behavior. Our understanding is based on the fact that direct conductance path between individual particles of different components ($CrO_2$ and $MgB_2$) is effectively blocked, and thus the maximum resistance of the composite corresponds to the maximum area of contact interface between the two types of particles, a phenomenon that becomes increasingly prominent at low temperatures. While this understanding is by no means complete and would require additional experimental and computational work to verify, it can predict the double percolation effect under certain conditions, consistent with our experimental observations. We expect that the results from this simple model system can be used to explore novel material properties of more complex multicomponent systems.

This work was supported by the National Science Foundation under NSF CAREER ECS-0239058, by DARPA through ONR Grant N00014-02-1-0886, and by ONR Grant N00014-06-1-0616.

FIG. 1: (Color online) (a) EDS image of a $CrO_2/MgB_2$ sample with $x = 0.4$ (just below the percolation threshold), showing the distribution of Mg (green) and Cr (blue). (b) SEM image of pure $CrO_2$ powder; the length of typical particle nanorods is about 300 nm, with approximately 8:1 aspect ratio. (c) SEM image of pure $MgB_2$ powder; the typical particle size is about 500 nm, with roughly spherical shape of particles.

FIG. 2: (Color online) Resistance of $(CrO_2)_x(MgB_2)_{1-x}$ samples as a function of weight fraction $x$ at different temperatures. Inset: Temperature dependence of the resistance for cold pressed samples of pure $CrO_2$ (top) and pure $MgB_2$ (bottom).

FIG. 3: (Color online) Scaling behavior of the resistance near the two percolative thresholds, at various sample temperatures as listed in Fig. 2: (a) the $CrO_2$ side of the transition, with $p_c^{CrO_2} = 0.34$ ($x_c^{CrO_2} = 0.5$); (b) the $MgB_2$ side, with $p_c^{MgB_2} = 0.78$ ($x_c^{MgB_2} = 0.65$).

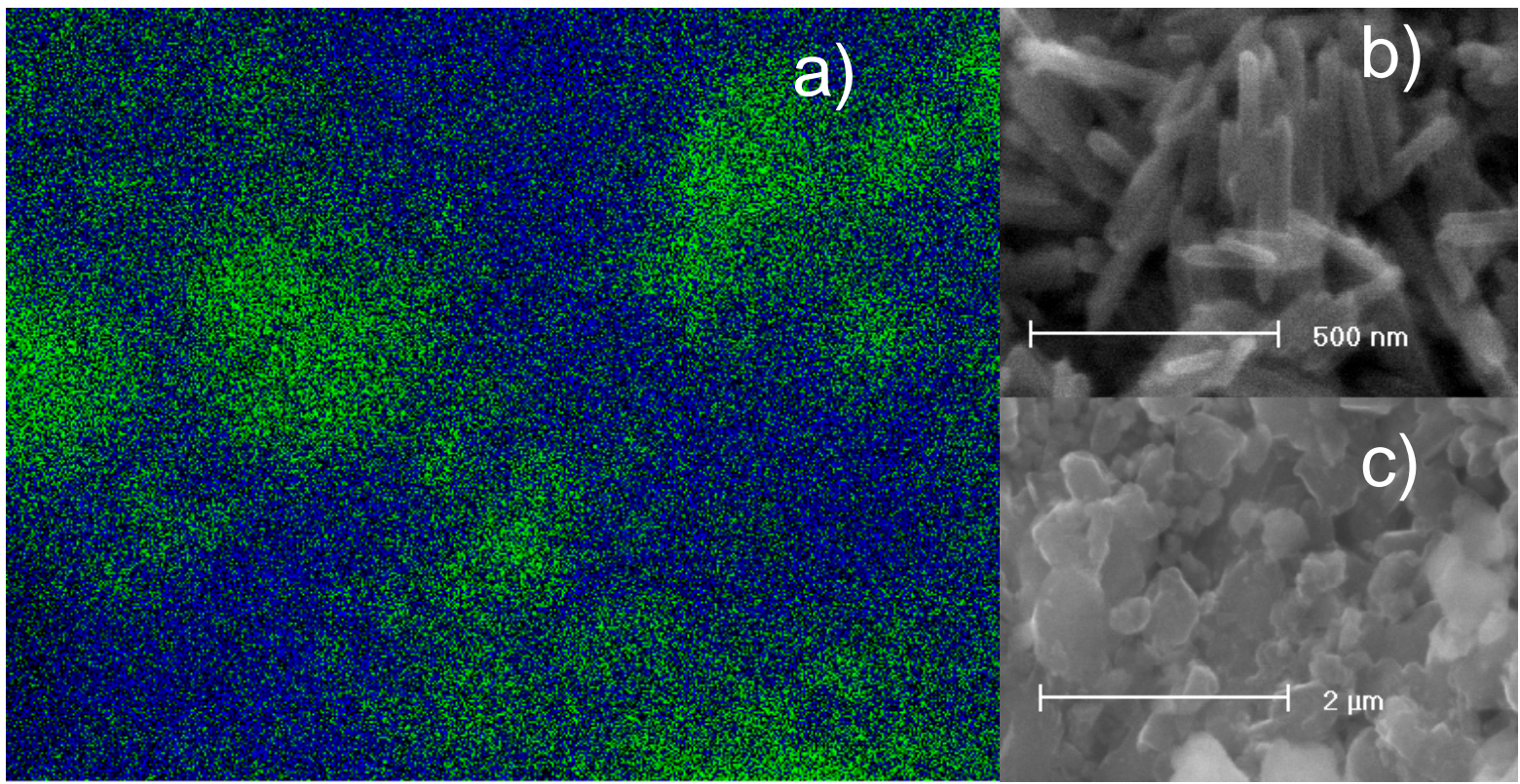

Figure 1 X. Liu, et al

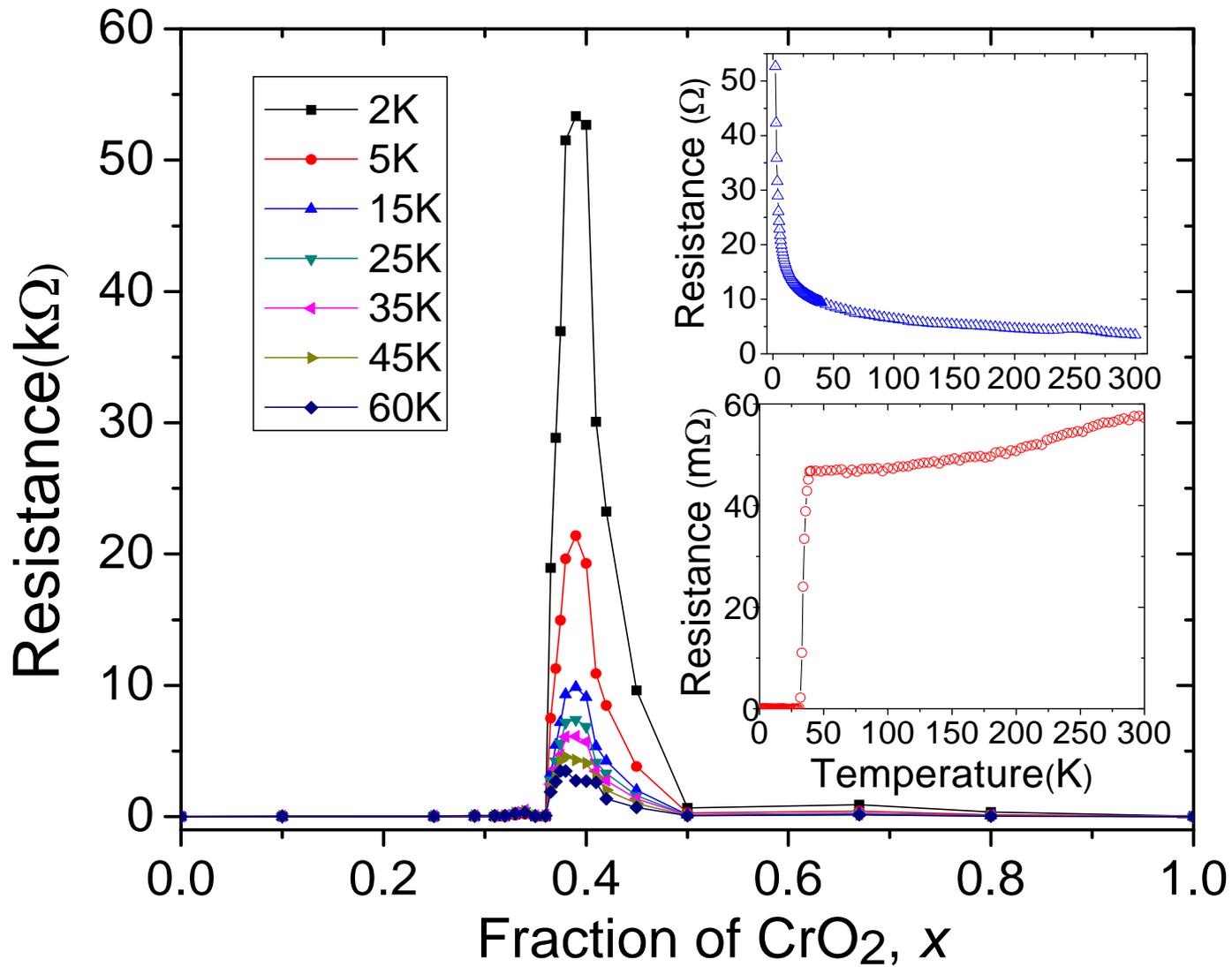

Figure 2  X. Liu, et al

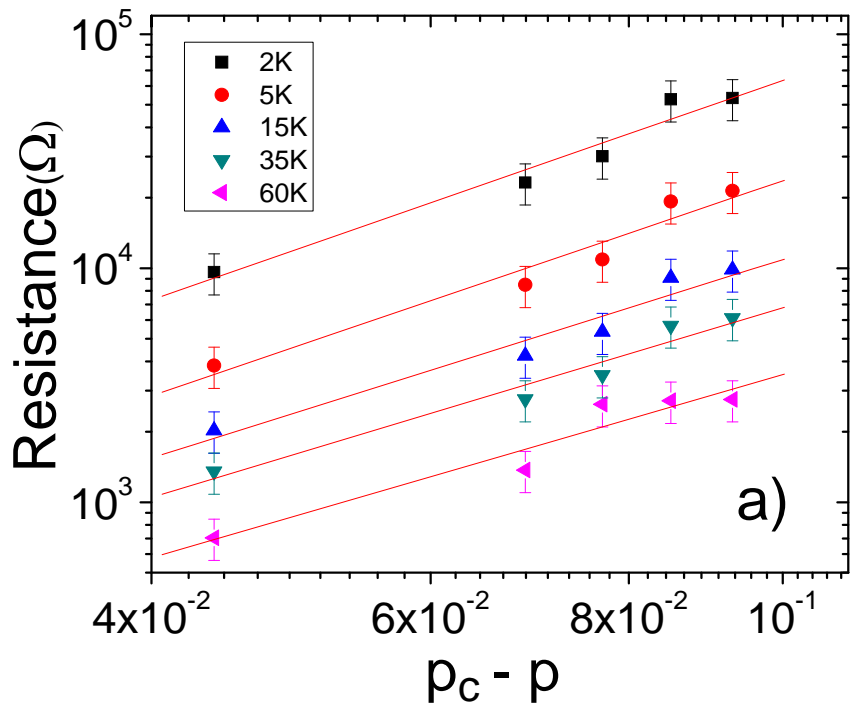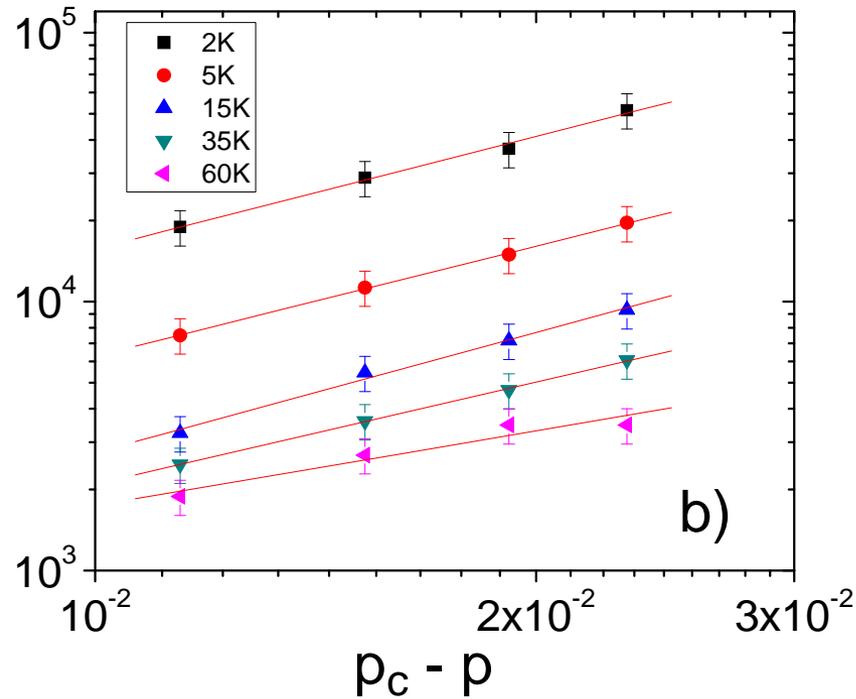

Figure 3        X. Liu, et al